# Overview and Applications of GPGPU Based Parallel Ant Colony Optimization

Sandeep U Mane[1], Pooja S. Lokare[2], Harsha R. Gaikwad[3]

[1,2,3] Dept of CSE, Rajarambapu Institute of Technology, Rajaramnagar, Sangli, MS, India

[1]*manesandip82@gmail.com*[1]
[2]*lokarepooja31@gmail.com*[2]
[3]*harshagaikwad4@gmail.com*[3]

**Abstract**

Ant Colony Optimization algorithm is a magnificent heuristics technique based on the behaviour of ants. Parallel computing is a mean to achieve the desired results in commensurable execution time. Parallelization of Ant Colony Optimization is utilized to solve large and complex problems. In this paper, review of different parallelization approaches for Ant Colony Optimization and its various applications discussed. Parallel Ant Colony Optimization has proved to be a successful approach for highly constrained problems in different areas such as routing, scheduling, timetabling etc. Parallelization of Ant Colony Optimization reduces the execution time; enable to increase the size of the problem, etc.

## 1. Introduction

Ant colony optimization (ACO) is a population-based search method inspired by the behavior of real ants. ACO was proposed by Dorigo and colleagues (2006) for solving hard combinatorial optimization problems.

Ants search for food, in the vicinity of their nest. During their mission, ants deposit a chemical substance called pheromone on the ground. The objectives of such activity are, to allow ants to find their way back to the nest and also it allows other ants to know the way they have taken, so that they can follow them (Dorigo and Stutzle, 2010). ACO has proved to be a successful approach for highly constrained problems in different areas. Marta et al. (2012) has provided a wide literature survey on ACO.

As ACO algorithm is population-based, it is a collection of agents working in team to find an optimal solution. ACO usually converge to near optimal solution in a certain number of iterations with considerable computational resources and time. To improve performance of ACO, parallel implementation is an excellent thought (Dorigo and Stutzle, 2001).

In this paper, section 2 outlines the working of ACO and GPGPU computing. Section 3 gives details about different approaches for parallelizing ACO found in literature. In section 4, few applications of parallel ant colony optimizations are discussed. Conclusion of study with future research direction is presented in section 5.

## 2. GPGPU Computing And Ant Colony Optimization

The GPGPU computing is advanced from the early 1980s to the late 1990s. The Fig.1 shows the CPU and GPU architecture. There is a large performance gap between many-threads GPUs and general-purpose multi-core CPUs. It is discussed in Kirk and Hwu (2010) as; design of CPU is optimized for sequential code performance. It makes use of sophisticated control logic to allow instructions from a single thread of execution to execute in parallel or even out of their sequential order while maintaining the appearance of sequential execution; large cache memories are provided to reduce the instruction and data access latencies of large complex applications (Kirk and Hwu, 2010).

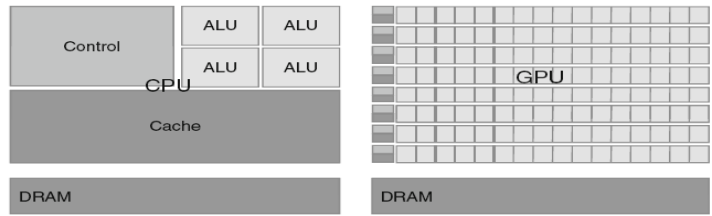

Fig.1. CPU and GPU Architecture

NVIDIA's CUDA architecture is a powerful platform for developing highly parallel and scalable applications. CUDA architecture provides simple abstractions for hierarchical thread organization, memories, and synchronization (Kirk and Hwu, 2010). CUDA architecture supports different programming languages, environments and libraries. Some of them are C, FORTRAN, OpenCL, OpenACC, Python and Thrust, etc.

The need of performance improvement in highly computational problems has made the GPGPU computing as emerging research area. GPGPU computing is an attraction of many researchers because of the low-cost, high availability and its SIMD (Single Instruction, Multiple Data) nature, which are inherently data-parallel.

In the CUDA framework, the GPU is viewed as a co-processor to the CPU. The GPU has its own device memory, and execute a very high number of threads in parallel. A kernel executes the data-parallel portions of an application on the GPGPU device using threads. CUDA organizes threads in execution into different blocks, which are mapped onto a multiprocessor in the GPU (NVIDIA, 2010). All the threads in one block are synchronized together and communicate with each other. The thread organization is as shown in Fig. 2 (Kirk and Hwu, 2010).

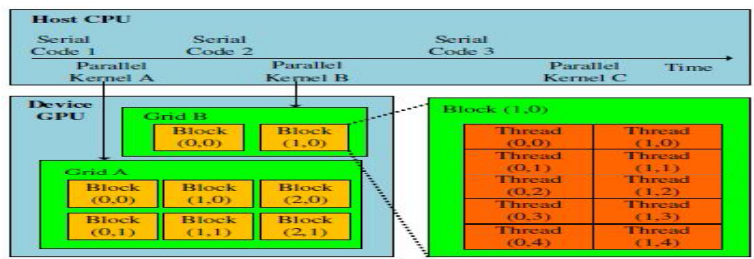

Fig.2. NVIDIA CUDA Architecture (Taken from Kirk and Hwu, 2010)

Threads within same block communicate each other and synchronized and the same kernel will be executed by them (Kirk and Hwu, 2010; Huang et al., 2008).

Dorigo et al. (2006) has stated that, Ant colony optimization (ACO) is a swarm intelligence population-based search method inspired by the social behavior of ant colonies. Ant communities have essential characteristics like, distributed collaboration, self-organization, adaptation, and distribution, which are useful to solve constrained or unconstrained optimization problems. Ant colony optimization algorithm is inspired by the pheromone trail laid by ants and their characteristic to use pheromones as a communication medium.

Scientist and researchers have successfully solved various combinatorial optimization problems using ACO algorithm and its variant like max-min ant system, ant colony system, rank based ant system, continuous orthogonal ant colony, etc. (Marta et al., 2012; Dorigo and Stutzle, 2010).

The ACO algorithm has four main steps, taken from Dorigo et al. (2006) and represented as:

*2.1. Initialization:*

Initial population is generated randomly, where initial population of the colony and the pheromone trail are set randomly. The starting node for all ants set randomly.

*2.2. Solution construction*

Ant selects the next node by observing the problem-dependent heuristic information and the trail intensity of the path. This step is repeated until all complete solution is constructed.

*2.3. Trail update*

Solution and deposited pheromone on the solution path are evaluated. If the greater amount of pheromone is deposited on the solution path, then better the solution.

*2.4. Pheromone evaporation*

The pheromone trail of all solution paths is decreased by some constant factor at the end of an iteration of building completed solutions.

**3. Different Approaches for Parallelizing Ant Colony Optimization**

Parallel implementations became popular with the intention of improving the efficiency of population-based search methods for solving complex optimization problems (Delisle 2012). In parallel implementations, the population is divided into several independent clusters and computing resources are assigned to process it. It benefits to heuristic techniques to obtain high quality results in a reasonable execution time, though optimization problems are hard-to solve (Dawson and Stewart, 2013). Parallel implementation of heuristic introduces a new exploration pattern useful to improve the result quality of the sequential implementations (Dorigo and Stutzle, 2010). The metric used to measure performance of parallel algorithms are speedup and efficiency (Randall and Lewis, 2009; Kothapalli et al., 2009). The speedup evaluates how much faster a parallel algorithm is than a corresponding sequential version.

To improve the results quality and the computational efficiency, in literature different approaches for parallel implementations of ACO are proposed.

To parallelize evolutionary algorithms, the traditional parallelization strategies are Master-slave model, cellular model, island model, and hybrid model. The master-slave model has subcategories as, Coarse-grain, Medium-grain and Fine-grain master-slave model. These models are used by researchers to parallelize ACO. Few other strategies are also proposed by researchers to parallelize ACO. In independent runs model communication between numerous sequential algorithms is absent. In multi-colony model, number of ant colonies is working independently but after some time interval colonies share information among them to achieve cooperation among colonies. The hybrid model is combination of other models (Randall and Lewis, 2009; Kothapalli et al., 2009; Li et al., 2011; Cecilia et al., 2011; Stutzle, 2001; Delevacq, 2013; Pedemonte et al., 2011).

The parallelization strategy to be used depends on the particular problem one has to solve and the available hardware. Most parallelization strategies can be classified into fine-grained and coarse-grained approaches. Characteristic of fine-grained approaches is that very few, often only one, individuals are assigned to one processors and individuals are connected by a population-structure. In coarse-grained approaches, larger subpopulations are assigned to single processors and information exchange is rather rare (Dorigo et al., 2006; Delevacq, 2013).

**4. Applications of Parallel Ant Colony Optimization on GPGPU**

The parallel ACO is used to solve problems in the combinatorial optimization, real-life routing and planning, industrial design. Bioinformatics, Telecommunications, other scientific problems, theoretical works.

Pedemonte et al. (2011) have provided extensive survey about applications domains, where parallel ACO models and its variants are successfully implemented to achieve better computational efficiency and quality results.

Uchida et al. (2012) proposed Ant colony optimization for solving the traveling salesman problem (TSP) on the GPU. Author has well thought–out programming issues. To choose next city randomly, computational cost effective iterative random trial method is proposed by Uchida et al., resulted in 42 times faster than conventional CPU implementation.

Caceres et al. (2012) used heuristic techniques based on ant colony system to find approximate solutions for the Quadratic Assignment Problem using GPGPU. The parallel version of Hybrid Ant System (HAS-QAP algorithm) and the cunning Ant System (cAS-QAP algorithm) are used to solve QAP problem.

Bai et al. (2009) introduced a Max-Min Ant System on GPU with CUDA.

Zhu and Curry (2009) implemented a coarse-grain master-slave ACO with a local search to solve bound-constrained continuous optimization problems using GPGPU.

Jiening and Jiankang (2009) proposed Max-min ant system on GPGPU to solve TSP. Cecilia et al. (2012) discussed several parallelization strategies for the ACO algorithm on the GPU. The data parallelism for ACO is enhanced using GPGPUs. Authors suggested that ACO is a potentially fruitful area for future research using GPU.

Islam et al. in (2003) proposed PACO on distributed environment using MPI to find shortest route in MANET.

Craus and Rudeanu (2004) proposed a generic framework to run evolutionary sequential algorithms in parallel environment. Author has tested ACO algorithm for TSP.

Liu et al. (2008) proposed multi-path routing protocol based on parallel ant colony optimization.

Hongwei and Yanhua (2009) proposed an improved Max-min ant system on MPI platform to solve DNA sequencing by hybridization problem, known as parallel improved Max-min ant system.

Guo et al. (2009) discussed 2D HP protein folding problem using parallel ACO. The authors have shared the pheromone matrix to integrate knowledge exposed by multiple ant colonies.

Yong et al. (2009) presented a parallel ACO with dynamic transition probability using MPI and C language to solve TSP. Dongdong et al. (2010) presented a multi-core parallel ACO using OpenMP and TBB –a multi-core computing technology for target assignment problem.

Green et al. (2011) implemented and compared population based intelligent search methods to improve the convergence time. Tsutsui and Fujimoto (2011) solved quadratic assignment problem by parallel ACO on GPGPU.

Day by day applications developed using parallel ACO and its variants on GPGPU, OpenMP or MPICH are increasing. Few applications developed using parallel ACO and its variants are summarized in Table 1.

Table1. Summary of Applications Using Parallel ACO & Its Variants

| ACO Type | Selected Problem/ Objective of Study | Author and Year |
| --- | --- | --- |
| PACO | Intelligent Motion Control for Omnidirectional Mobile Robots | Huang et al., 2011 |
| A Shared-Memory ACO | Numerical Optimization | Korosec et al., 2011 |
| GACO Tool using ACO | Design Validation | Li et al., 2011 |
| GA, ACO, PSO | Traffic Signal Coordination | Wangl and Shen, 2012 |
| ACS | Quadratic Assignment Problem | Caceres et al., 2012 |
| SoPC-Based PACO | Optimal Motion Controller Design for Mobile Robots | Huang, 2013 |
| ACO on OpenMP & CUDA | Shortest Path Problem in Vehicle routing | Arnautovic et al, 2013 |
| ACO & OpenMP | TSP | Milanovic et al., 2013 |
| ACO on CUDA | Path Planning for Unmanned Aerial Vehicles | Cekmez et al., 2014 |

## 5. Conclusion And Research Scope

Ant Colony Optimization (ACO) belongs to the family of population-based search method that has been successfully applied to many NP-hard problems. As a population based algorithm, it is intrinsically parallel, and thus well suited for implementation on parallel architectures.

This paper, gives review of parallel ACO with different applications of Parallel ACO. Parallel implementations of heuristic techniques allow reaching high quality results in a reasonable execution time. The scalability of the GPU is excellent for parallel problems such as optimization problems. Overall, this shows that it is possible to significantly reduce the execution time of ACO on GPU. Though, the GPU has improved significantly both in terms of programmability and performance, It performs particularly well on data parallel applications. Thus an argument can be made that, a GPU as a whole is a good candidate for general purpose computing.

GPGPU is good option to improve speedup while solving optimization problems using Ant Colony Optimization. The speed-up obtainable through parallelization should be highly important for solving larger problem instances in real time. In future the constraint satisfaction problems, employee scheduling problems can be solved with parallelization of ACO on GPGPU.